\documentclass{aastex63}
\usepackage{amsmath}
\usepackage{multirow}

\graphicspath{{./}{figures/}}

\makeatletter
\g@addto@macro\@floatboxreset\centering
\makeatother

\begin{document}

\title{Searching for Binary Asteroids in Pan-STARRS1 Archival Images}

\correspondingauthor{James Ou}
\email{james82@hawaii.edu}

\author[0000-0002-8439-7767]{James Ou}
\affiliation{Institute for Astronomy, University of Hawai`i at M\={a}noa, Hilo, HI 96720, USA}

\author[0000-0002-1917-9157]{Christoph Baranec}
\affiliation{Institute for Astronomy, University of Hawai`i at M\={a}noa, Hilo, HI 96720, USA}

\author[0000-0003-4191-6536]{Schelte J. Bus}
\affiliation{Institute for Astronomy, University of Hawai`i at M\={a}noa, Hilo, HI 96720, USA}

\begin{abstract}
We developed two different point spread function (PSF) analysis techniques for discovering wide separation binary asteroids in wide field surveys. We then applied these techniques to images of main belt asteroids in the 4 to 60 km size range captured by Pan-STARRS1. \citet{Johnston2019} lists fewer than 10 known binaries in this size range with separations greater than 10\% of the primary's Hill radius, so discovering more wide binary asteroids is crucial for understanding the limits of binary stability and improving our knowledge of asteroid masses. We analyzed each image by: \emph{i}) comparing the major axis orientation of the asteroid’s elliptical PSF to its non-sidereal rate on the sky, and \emph{ii}) comparing the one-dimensional median profile created by collapsing the image along the asteroid’s direction of motion to that of nearby field stars. For both methods, we flagged any results that deviated significantly from the expected measurements of single asteroids, and those targets with the most flags were identified as binary candidates for confirmation with high-acuity imaging.

\end{abstract}

\section{Introduction}
Mass is a fundamental characteristic of asteroids, influencing how they form, evolve, and interact with other objects. Large scale astronomical surveys have provided abundant information on other asteroid characteristics such as color and size; however, the determination of masses has proved much more elusive.  While it is possible to infer asteroid masses from their spectral colors, albedo and size, combined with knowledge from meteorite analogs, assumptions about their dynamical histories must also be included. Rubble pile asteroids are believed to form through the collisional disruption of parent bodies followed by the reaccretion of material, and the resulting bodies have porosities dependent on the physical conditions. After their formation, asteroids may undergo deformation over time as they are exposed to a variety of effects; 162173 Ryugu, visited by Hayabusa2, has a shape that may have been caused by periods of high spin rate (\citealp{Watanabe2019}). Furthermore, laboratory experiments have shown that individual impacts where the debris collapses back on the parent may cause considerable bulking, and if the body already has high macroporosity, then shocks acting as compaction waves may cause a net reduction (\citealp{Britt2002}). However, simulations of asteroid dynamical formation and evolution are limited by how porosity is treated in the calculations\footnote{In order to improve computational efficiency, particles in contact after the initial reaccretion phase of asteroid formation simulations are merged into superparticles with an assumed constant bulk density (e.g., \citealp{Michel2004}).}. Asteroid masses have been estimated by measuring and modeling orbital perturbations caused by other solar system bodies with known masses, though this technique is limited by the precision of measurements, e.g., Gaia DR2 milliarcsecond astrometry for $\sim$100 km main belt asteroids (\citealp{Siltala2022}). There may be a correlation between size and macroporosity, but a number of cases exist where measurements of asteroids of similar size and type result in very different results (e.g., $11\pm19\%$ for 93 Minerva vs $61\pm7\%$ for 90 Antiope, \citet{Carry2012}). If bulk density or mass is not measured and is instead assumed, then the uncertainty from such a wide potential range of macroporosities propagates through estimations of other asteroid characteristics. This leads to greater ambiguity of inferred properties, such as internal structure and tensile strength, which in turn limits our understanding of their formation, evolution, and lifetime.

The existence of binary asteroids provides a direct way for determining the precise masses of those systems by monitoring how the two components orbit about their common barycenter (\citealp{Carry2012}). The first confirmed binary asteroid, 243 Ida and its satellite Dactyl, was discovered through a close approach by the Galileo spacecraft in 1993 (\citealp{Chapman1996}) and we now know of over 400 minor planets with natural satellites. These have been discovered through a number of methods: analyzing photometric light curves is so far the most productive technique for identifying natural satellites around minor planets, accounting for over $\sim$50\% of discoveries; resolving companions through diffraction-limited imaging accounts for $\sim$35\% of discoveries; and radar observations account for much of the remainder, excepting extraordinary events such as stellar occultations and spacecraft flybys (\citealp{Johnston2019}). 

Each of these methods is suited for a different region of the contrast, angular separation, and distance parameter space, with very different observational opportunity costs. Light curve analyses are highly efficient through a variety of completed, on-going, and future large scale time-domain photometric surveys; their discoveries require an eclipsing orbital plane orientation that favor binaries with smaller separations, and must consider inherent variations from non-spherical asteroid shapes (e.g., \citealp{Szabo2017}). \citet{Warner2018} demonstrated that very wide binary asteroids, where the companion is not tidally locked, can be identified through light curves, but such objects are very rare. While diffraction-limited imaging is observationally more costly, requiring time on the heavily oversubscribed Hubble Space Telescope or with adaptive optics on large ground-based telescopes, it has the advantage of detecting companions over a wider range of separations and at much greater contrast ratios (e.g., \citealp{Marchis2006}). It is also robust to the orientation of the orbital plane. Radar provides more complete knowledge across both the separation and contrast ratio domains, but the power received $P_r$ from a reflected radar signal falls off rapidly with distance, $P_r\propto r^{-4}$, limiting radar studies to those asteroids whose orbits bring them closer to the Earth (\citealp{Naidu2016}).

\begin{figure}
\includegraphics[scale=0.95]{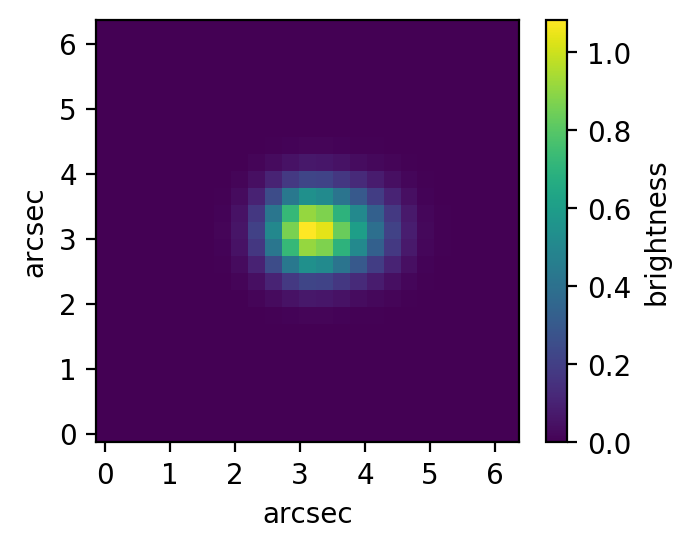}
\includegraphics[scale=0.95]{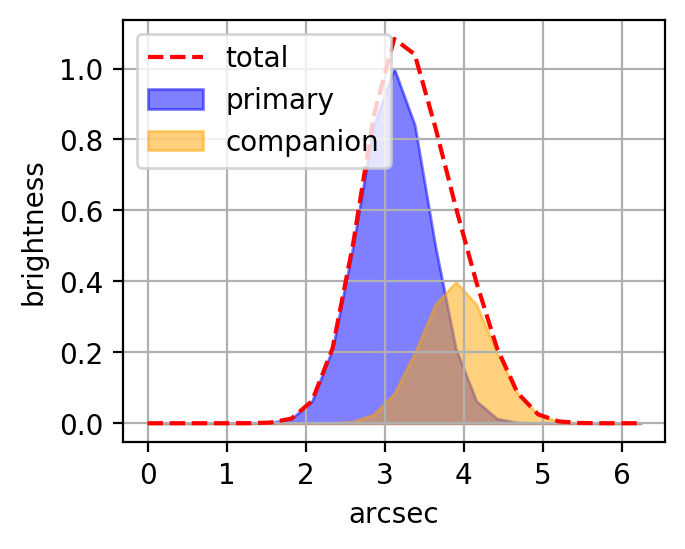}
\caption{These figures show a simulated PS1 image (0\farcs26 pixel scale) and its corresponding cross section, of a binary object with a 1 mag contrast ratio, 1\farcs04 FWHM, and 0\farcs78 apparent separation. While the companion is not resolved, its presence is evident from the asymmetric elongation of the combined profile.\label{fig:PSFs}}
\end{figure}

Here, we present two techniques to identify possible binary asteroids in the vast quantity of data available through ongoing and upcoming large scale, wide field imaging surveys. In images, unresolved binaries can produce a measurable difference in the shape of their point-spread functions (PSFs, e.g., fig. \ref{fig:PSFs}) and these techniques aim to identify and quantify these differences. \citet{Terziev2013} demonstrated that measurements of ellipticity could be used to identify binary stars with separations down to $\sim1/5$ of the seeing limit in the Palomar Transient Factory wide-field survey data, and \citet{Deacon2017} showed similar performance using images from the Panoramic Survey Telescope and Rapid Response System (Pan-STARRS), following the algorithm described by \citet{Hoekstra2005}. Binary asteroids differ from binary stars in that companion orbits that would measurably distort wield-field survey PSFs have periods on the order of days, and their PSFs have inherent shape variance due to their non-sidereal trailing on the sky plane during the exposures. While these two aspects prevent us from using the exact techniques described by Terziev and Deacon, we attempt similar analyses that account for these differences: (1) measurement of the direction of the PSF major axis, which we expect will vary in conjunction with a companion's orbit, and (2) the PSF marginalized along the direction of the ephemeris motion, where asymmetries of the profile may reveal the presence of a companion. Once binary asteroid candidates have been identified, follow-up high-acuity imaging will still be necessary for confirmation, refining satellite orbits, and determining system masses. Pre-screening with the wide field imaging data reduces the observational cost of searching for binaries, both for increasing the sample size of asteroids with precise mass and porosity estimates, and to probe the properties of companion populations near outer separation limits.

\section{The Main Belt Binary Asteroid Population}
The majority of known asteroids reside in the Main Belt, where the number of objects at least 1 km in diameter is on the order of $10^6$ (\citealt{Tedesco2002}). In this region, binary asteroids are expected as a product of asteroid family formation. The catastrophic collisional disruption of parent bodies produces debris, which then reaccretes into a population of family members with similar proper orbital elements, composition, and weathering/dynamical age (\citealp{Cellino2002}). During this process, debris can enter orbit around what remains of a parent, or debris on similar trajectories can become bound to each other, such that the reaccreted bodies form satellites. An analysis by \citet{Nesvorny2015} found family memberships for 37\% of the main belt population that had been numbered at the time. In the formation of collisional families, the initial mass of the parent body, impact velocity, and incident angle all contribute to the number of binary systems and type of companions formed (\citealp{Durda2004}).

Other mechanisms of binary formation include tidal disruption (e.g., due to close approaches with planets) and fission when spun up to critical levels of angular momentum (e.g., due to the YORP effect), though these are more likely for asteroids in the inner Solar System. The fraction of binaries among NEAs is $\sim15\%$ (\citealp{Pravec2007}), and photometric light curve discovery rates in the main belt have been consistent with this fraction (\citealp{Margot2015}), despite the different expected formation mechanisms.

\begin{figure}
\includegraphics[scale=0.95]{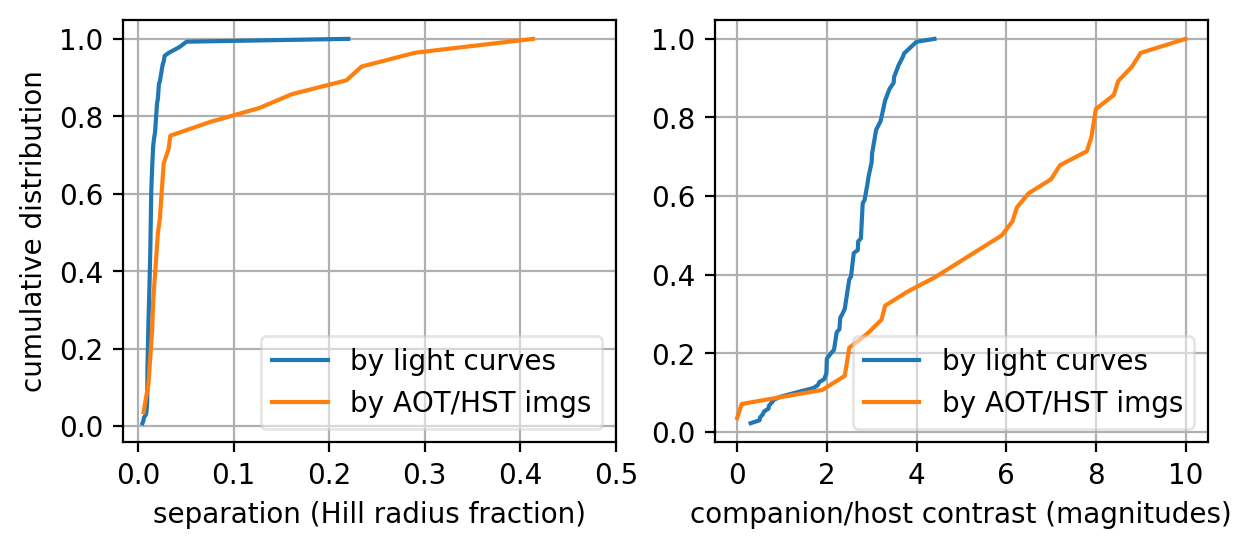}
\caption{Main belt binary asteroid separations and contrast ratio distributions produced from 137 companions detected by light curves and 28 companions resolved by direct imaging (\citealp{Johnston2019}). Direct imaging is better able to explore the binary population where separations are greater than $\sim0.1$ Hill radius, or where contrast ratios are greater than $\sim4$ magnitudes.\label{fig:known_binaries}}
\end{figure}

A compilation of known main belt binary asteroids (\citealp{Johnston2019}) provides us with 137 companions detected by analysis of light curves and 28 detected by diffraction limited imaging. As listed, the companions that were discovered by light curves have separations of $\lesssim50$ km and contrast ratios $\lesssim4$ mag, whereas those detected by direct imaging have separations from 84 km to 3340 km and contrast ratios as large as $\sim10$ mag. Cumulative distributions for the separations and contrast ratios of known main belt companions are shown in figure \ref{fig:known_binaries}. Notably, there is an absence of companions beyond $\sim$0.5 Hill radii; it may be that such orbits are unstable in the main belt environment, or that wide companions are unlikely to be produced through the mechanisms of binary formation.

\section{Data}
\subsection{Target Selection}
To select targets for analysis in wide field data, we take the following properties into account: potential separations, brightness, pixel sampling, seeing, and non-sidereal rate.

When observed on the plane of the sky, the angular separation between members of binary systems vary over time. Because the resulting distribution from the perspective of the observer favors the wider end of the range, we assume a geometry where the host-companion axis is at right-angles to the observer's line of sight.

If a companion exists, we expect it to occur within the region of stable satellite orbits delimited by the host's Hill radius $r_H$, which corresponds to the host's mass, $m\propto r_H^3$, and in turn to the size of its reflecting surface, $s\propto m^{2/3}$, and thus brightness. We define the angular separation space where we can reasonably expect to find companions by estimating the Hill radius and brightness at opposition from representative asteroid properties (see table \ref{tab:type_prop}) for specified spectral types, sizes, and semi-major axes. We do not use JPL Horizons brightness calculations, as we wish to normalize all values delimiting the angular separation space at zero phase angle and specified semi-major axes. For densities, we make use of the improved estimates from \citet{Vernazza2021} for 42 large main belt asteroids, which take into account 3D shape reconstruction and analyses. While larger asteroids trend towards lower porosities, resulting in higher estimated masses and thus larger Hill radii for a given host diameter, this is reasonable for the purpose of estimating outer bounds. Apparent brightness is calculated from heliocentric distance $r$, geocentric distance $\Delta$, diameter $d$, and geometric albedo $p_{V}$ using Ceres as a reference point:

\begin{equation}
\mathrm{mag}_{V}=\mathrm{mag}_{H,Ceres}+5\log\left(r\Delta\right)-5\log\left(\frac{d}{d_{Ceres}}\right)-2.5\log\left(\frac{p_{V}}{p_{V,Ceres}}\right) 
\end{equation}

\begin{table}
\caption{Representative Asteroid Physical Properties by Spectral Type (\citealp{Vernazza2021}, \citealp{Warner2019}). \label{tab:type_prop}}
\begin{tabular}{|c|c|c|}
\hline 
Property & S type & C type\tabularnewline
\hline 
\hline 
Density (g/cm\textsuperscript{3}) & 3.0 & 1.8\tabularnewline
\hline 
Albedo & 0.20 & 0.06\tabularnewline
\hline 
\end{tabular}
\end{table}

\begin{figure}
\includegraphics[scale=0.95]{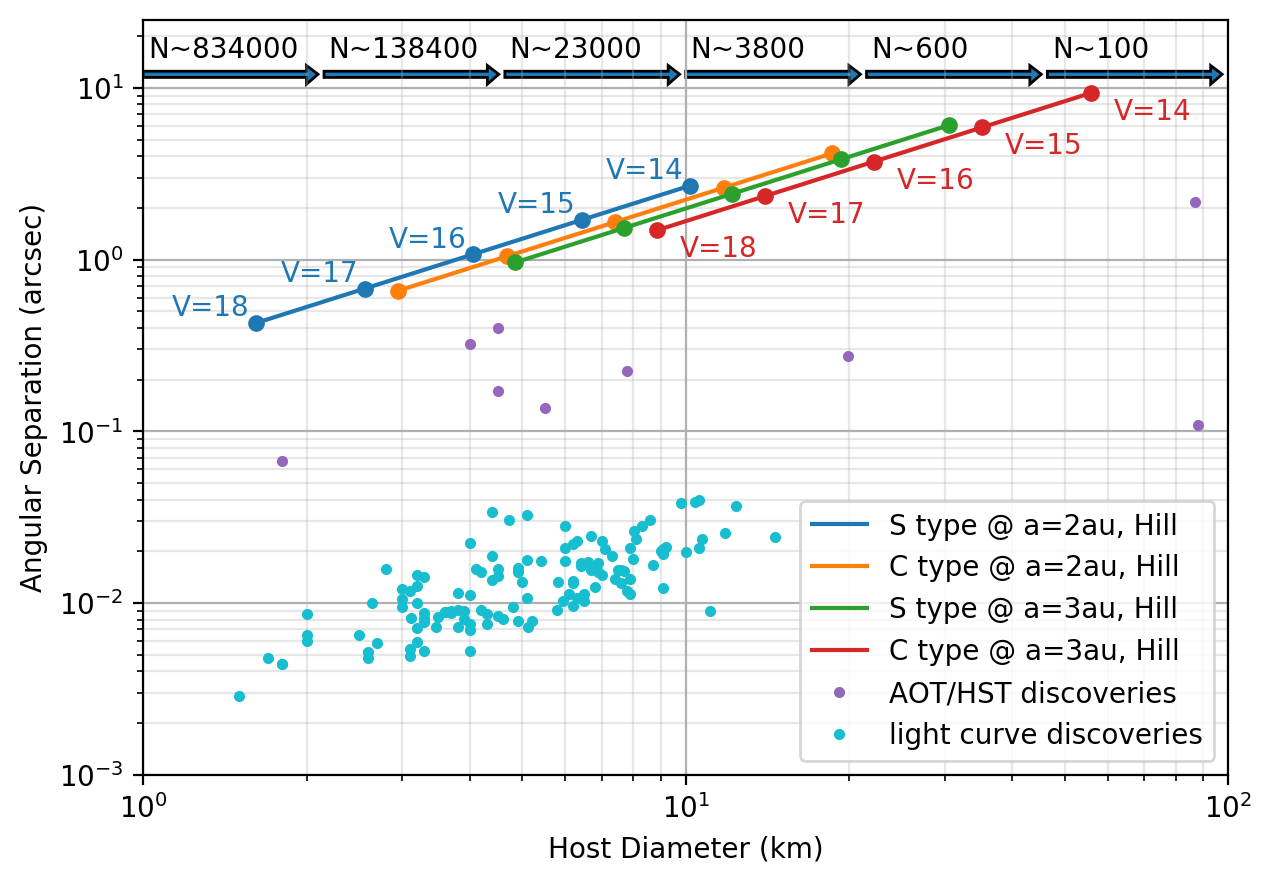}
\caption{The lines show the theoretical maximum possible host-companion angular separations (i.e., Hill radii) for representative S and C type asteroids located in the inner and outer main belt, observable at ${m}_V$ = 14-18 at opposition, corresponding to asteroid size. At the top are an estimate of the number of main belt objects following the size-frequency power law distribution from \citet{Ryan2015} and scaled to a total population of 1 million. The known binary asteroids show the different parameter spaces favored by each method of discovery. Better sampling in the direct imaging parameter space is needed to understand the real distribution of companion separations, which will provide constraints to main belt companion orbital stability and/or formation and evolution mechanisms.\label{fig:asep_v_diam}}
\end{figure}

These relationships for model S and C type asteroids at the inner ($a=2$ au) and outer ($a=3$ au) main belt are shown in figure \ref{fig:asep_v_diam}, along with the known main belt binaries of size $<100$ km. For the known main belt binaries, we use the estimated diameters and masses as compiled by \citet{Johnston2019}. If it is the case that real separations are limited to $~0.5$ Hill radii, then it would be reasonable for observable separation distributions to extend to $\sim1$ arcsec for 10 km diameter asteroids.

While the techniques we developed here can be applied to datasets from a variety of surveys, there exist a number of technical and physical limitations. For the purposes of a proof of concept here, we want to pair the telescope to a target asteroid population, such that it is reasonable to expect host-companion separations to be at least Nyquist sampled. It is possible to reduce this limit with more information: both Terziev and Deacon demonstrated the ability to detect binary stars with $<2$ pixel separations from the shape of the PSF by using multiple observations of each target.

Beyond the imaging system, we must consider that ground-based observations are limited in their acuity by atmospheric seeing. The ideal target population should have stable satellite orbital regions corresponding to separations that are not too much smaller than the seeing.

Finally, we must consider the limitations of the technique with regard to the method of observation. Large scale, wide field surveys are conducted using sidereal tracking, whereas asteroids trail at non-sidereal rates defined by their heliocentric orbits. Targeting asteroids with methods that involve the shape of the PSF must account for this trailing over the length of the exposure times. If the magnitude of trailing is too large, it will dominate measurements and limit the detectability of possible asteroid companions.

When taking these considerations into account, we determined that the Pan-STARRS surveys are suitable for searches of binary asteroids in the main asteroid belt. In 45 second PS1 exposures, inner belt asteroids ($a\sim2$ au, assuming circular orbits) at opposition have linear non-sidereal trailing of 2 pixels ($0\farcs54$) and outer belt asteroids ($a\sim3$ au) at opposition have trailing of 1.5 pixels ($0\farcs39$). To ensure that our dataset includes observations of targets which are sufficiently bright, but not saturated, we selected asteroids which reach brightness between $m_v\sim$ 14 and 16 from main belt asteroid families given by \citet{Nesvorny2015}. The limit of $m_v = 16$ is also typical for AO systems (e.g., Wizinowich2013), which will be needed to follow-up on identified binary candidates. This results in a list of 1,929 target asteroids.

\subsection{Pan-STARRS1}
Pan-STARRS is a wide-field astronomical imaging facility located on Haleakala, Hawaii. Pan-STARRS1 (PS1) is the first telescope of the facility, comprising a 1.8 meter telescope and 1.4 gigapixel GPC1 camera, with a 0\farcs26 pixel scale and a 7 square degree field of view. The available asteroid imagery come from a variety of PS1 observation programs from 2009 through 2019, using the $g$, $r$, $i$, $z$, $y_{PS1}$, and $w_{PS1}$ filters. Median seeing by filter vary from $1\farcs31$ to $1\farcs02$, with saturation typically occurring between $m_V$ = 12 and 14 depending on seeing, filter, and exposure time (\citealp{Magnier2013}). PS1 data releases provide ``warp" images that have been resampled into RINGS.V3 tessellation sky cells and are astrometrically and photometrically calibrated (\citealp{Waters2020}).

We obtained a total of 510,000 warp images from the PS1 data archive, each of which are $1400\times1400$ pixel ($6\times6$ arcmin) cutouts centered on the target asteroid. Of these, 106,000 satisfied our image viability requirements:
\begin{enumerate}
  \item Target aperture pixel values are in the range of linearity\footnote{\citet{Chambers2016} reports GPC1 non-linearity at 40,000 DN; we used a 20,000 DN threshold out of an abundance of caution.}.
  \item Target aperture does not contain masked pixels.
  \item Image contains at least 10 field stars that satisfy the above two requirements.
\end{enumerate}
The third requirement enables production of model PSFs for each image that are used to estimate image PSF full-width at half-maximum (FWHM) and to measure systematic PSF distortion. The FWHM for each image was estimated by taking the profiles of the field stars, measuring the separation in each profile where pixels have DNs at half of the peak, and determining the median. Of the images that meet these requirements, 80\% have 45 second exposure times, 12\% have 30 second exposure times, and the remainder have exposure times ranging from 3.6 to 600 seconds.

\section{PSF Major Axis Orientation}

\begin{figure}
\includegraphics[scale=0.95]{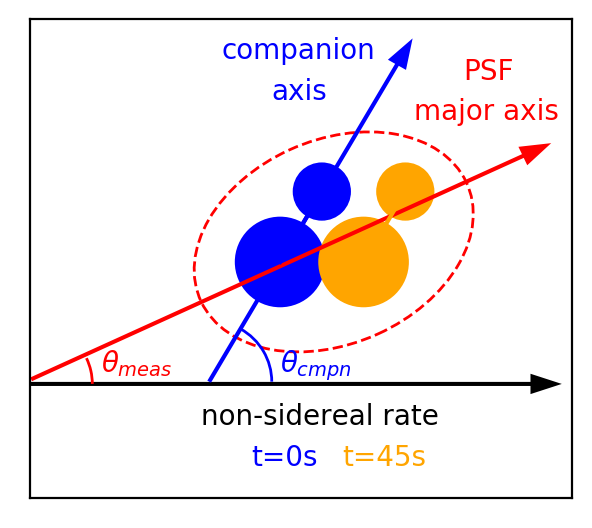}
\caption{In sidereal tracked observations, a binary asteroid's PSF is a product of both its members and its non-sidereal rate over the exposure time. The direction of its ellipse major axis may deviate substantially from the known non-sidereal rate depending on the direction, separation, and brightness of the companion.\label{fig:major_axis}}
\end{figure}

\subsection{Concept}
On the two dimensional plane of an image, the PSF of a target can be modelled as an ellipse, with a major axis along its longest diameter. A non-aberrated PSF of a point source, from a circular/annular aperture, is perfectly round with an ellipticity of zero. If the target has a non-sidereal motion and is observed with sidereal tracking, then there is an inherent elongation producing an ellipse major axis aligned with the direction of its non-sidereal rate. If there is a second source, such as a companion, then the combined PSF may produce a major axis that deviates from the non-sidereal rate (e.g., see fig. \ref{fig:major_axis}), and that is what we are attempting to measure with this technique.

The ellipticity of a PSF can be characterized through central image moments $I$, e.g., \citet{Kaiser1995}, \citet{Hu1962}:

\begin{equation}
\begin{aligned}I_{11}= & \sum_{x,y}x^{2}f\left(x,y\right)W\left(x,y\right)\\
I_{22}= & \sum_{x,y}y^{2}f\left(x,y\right)W\left(x,y\right)\\
I_{12}= & \sum_{x,y}xyf\left(x,y\right)W\left(x,y\right)
\end{aligned}
\end{equation}

In order to prevent our calculations from being dominated by photon noise as we move farther from the PSF centroid, it is necessary to use a weight function $W\left(x,y\right)$. Similar to \citet{Deacon2017}, we use a circular Gaussian with the same FWHM as the seeing. From these moments, we calculate two ellipticity parameters which describe the vertical/horizontal and diagonal polarizations, followed by the ellipticity $e_{tot}$ and major axis orientation angle $\theta$:

\begin{equation}
\begin{aligned}e_{1}= & \frac{I_{11}-I_{22}}{I_{11}+I_{22}}\qquad e_{2}=\frac{2I_{12}}{I_{11}+I_{22}}\end{aligned}
\end{equation}

\begin{equation}
\begin{alignedat}{1}e_{tot}= & \sqrt{e_{1}^{2}+e_{2}^{2}}\qquad\theta=\frac{1}{2}\arctan\left(\frac{e_{2}}{e_{1}}\right)\end{alignedat}
\end{equation}

The calculated angle $\theta$ is relative to the nearest Cartesian axis defined by the image's pixel grid, and the signs of other central image moments can be used to identify which axis. Since all of our targets have precise non-sidereal rates available through the NASA/Jet Propulsion Laboratory Horizons ephemeris computation service (\url{https://ssd.jpl.nasa.gov/horizons/}), we measure deviations from this known source of ellipticity in the PSF. Non-zero residual angles provide possible evidence for the presence of a companion. 

In practice, there are several error sources which affect our measurements: photon noise; sampling noise due to angular measurements of square pixel data and sub-pixel positioning of the PSF on the pixel grid (e.g. where the peak is centered on a pixel vs at an edge); contamination by other astronomical objects; and optical distortion from the atmosphere, instrument, and telescope. Photon and sampling noise are estimated through modelling, and source blends can be handled by identifying objects in the region while the asteroid is elsewhere.

Solving the optical distortion can be complex. Normalizing PSF variance due to seeing requires multiple observations with the same host-companion positioning over different atmospheric conditions (\citealp{Terziev2013}), but the short orbital period of asteroid companions, on the order of days, requires a much higher cadence than is available. Correcting for PSF anisotropy across a detector requires some number of well distributed point sources to fit (\citealp{Hoekstra2005}), and any remaining anisotropy also requires multiple observations to reduce uncertainty, again with the same observing geometry. With the large number of available images, a simple solution is to only use those images where the PSFs of nearby point sources have ellipticities constrained below selected thresholds that would indicate good observing conditions.

\subsection{Simulations}
\begin{figure}
\includegraphics[scale=0.95]{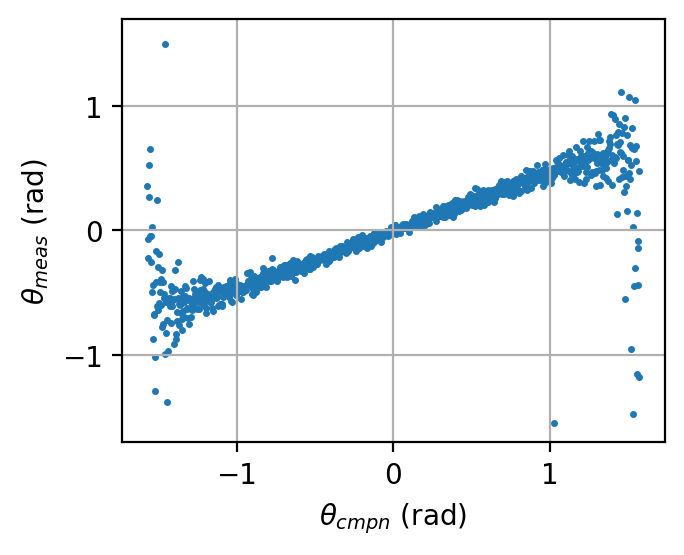}
\caption{Simulations show that it is possible to detect companions at low contrast ratios despite a stronger and brighter effect from non-sidereal trailing. Here, the binary system has a contrast ratio of 1 mag, separation of 1.5 pixel (0\farcs39), and non-sidereal trailing of 2 pixels (0\farcs52); companion positioning at $\pm1$ rad produce major axis orientation measurements of $\pm0.5$ rad. \label{fig:companion_theta}}
\end{figure}

To estimate the variance in major axis orientation for solitary vs. binary asteroids, we conducted a number of Monte Carlo simulations using Gaussian PSFs with parameters matching those expected in the PS1 imagery: peak pixel brightness of 1000 to 5000 DN, FWHM of 1\arcsec to 2\arcsec, and Poisson pixel noise (with 1.3 $e^-$/DN gain and 2000 DN background). Simulations of binary asteroids with a 3 magnitude contrast ratio reliably produce measurements different from simulations of solitary asteroids when both non-sidereal trailing and separations exceed 1.5 pixels (0\farcs39; fig. \ref{fig:companion_theta}).

\begin{figure}
\includegraphics[scale=0.95]{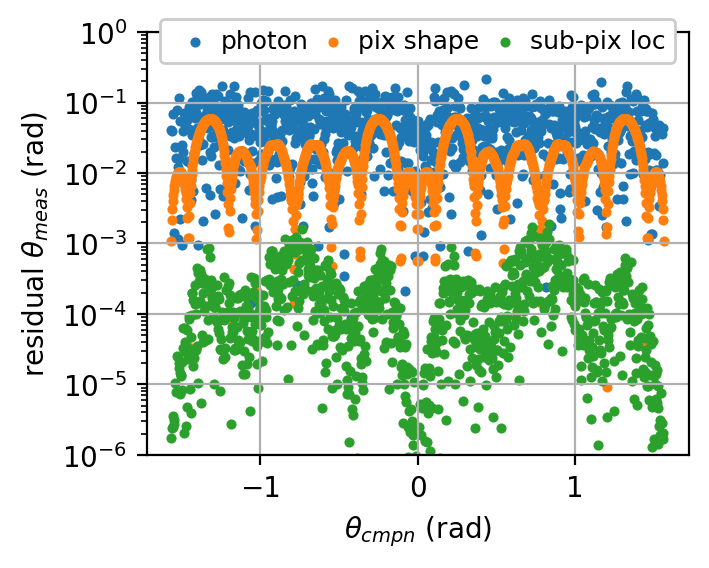}
\caption{Sources of major axis orientation measurement errors include photon noise, angular measurements of square pixel data, and the sub-pixel positioning of the PSF on the pixel grid. Simulations of a point source show that the square pixels only significantly affect measurements at certain angles, and that sub-pixel PSF positioning has negligible effect when the PSF FWHM is similar to typical PS1 imagery.  \label{fig:meas-components}}
\end{figure}

From our simulations, we also found that photon noise is the greatest source of error despite using the circular Gaussian weight function to reduce its influence (see fig. \ref{fig:meas-components}). Measuring PSF major axis directions that are not aligned with the axes of the square pixel grid can approach similar contributions to noise, depending on the angle at which the non-sidereal trailing was placed. The effect of sub-pixel positioning of the PSF, i.e., whether the peak is centered on a pixel, or on an edge, or somewhere in between, is an order of magnitude less than the photon noise at FWHM = 3 pix and three orders of magnitude less at FWHM = 8 pix, so we consider it negligible for processing of the PS1 data.

For our solitary asteroid baselines, we conducted a series of simulations each consisting of 1 million iterations. Their non-sidereal motions were randomized, with specified seeing and brightness thresholds, and photon noise generated from the brightness and background. We then binned the results, such that low count bins identify regions where measurements of singular asteroids are least likely to fall. Because ellipses have two axes of symmetry, it is only necessary to simulate from $0$ to $\pi/2$. 

Different baselines were produced for PSF amplitudes of 5000, 4000, 3000, and 2000 DN, and seeing FWHM from 4 to 8 pixels. Simulations with dimmer PSFs and worse seeing did not have distinct outlier regions for singular asteroids with non-sidereal rates appropriate for the main belt. Very few main belt asteroids are likely to have been observed with peak pixel brightness greater than 5000 DN or with better seeing than 4 pixels, so we did not produce baselines at brighter PSFs or better seeing.

We also simulated asteroids for the upcoming Legacy Survey of Space and Time (LSST) at the Vera C. Rubin Observatory, considering a 90,000 $e^-$ CCD well, $0\farcs7$ FWHM, and $0\farcs20$ pixel scale. We find that we should be sensitive to binaries when both trailing and separations exceed 1.3 pixels ($0\farcs26$). However, the 15 second exposures of LSST's baseline survey will only produce sub-pixel trailing of main belt asteroids ($0\farcs18$ at $a\sim2$ au, $0\farcs13$ at $a\sim3$ au); too little trailing greatly increases the uncertainty for the expected values of single asteroids, reducing the reliability of this technique.

\section{Cross-track Profiles}

\begin{figure}
\includegraphics[scale=0.95]{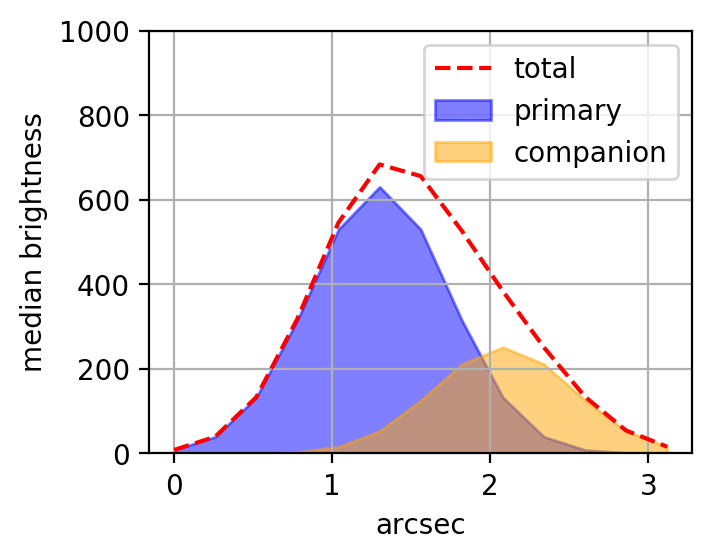}
\includegraphics[scale=0.95]{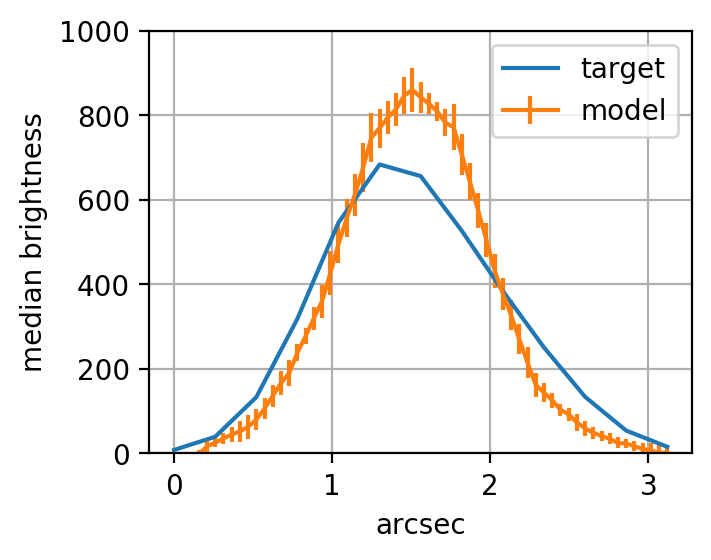}
\caption{The cross-track profile of a target asteroid is created by collapsing its PSF in the direction of its non-sidereal motion (left) and is compared to a model profile created from the field stars in the same image, normalized to the same total brightness (right). The deviation between this simulated binary (1 mag contrast ratio) and the corresponding model profile is large, and can be quantified with a goodness-of-fit statistic. \label{fig:crosstrack}}
\end{figure}

This technique allows for the comparison of one-dimensional profiles of the target asteroid with model produced from field stars (see fig. \ref{fig:crosstrack}), regardless of PSF asymmetries. By collapsing the PSF in the direction of the target asteroid's non-sidereal motion, we produce marginal profiles where the trailing from sidereal tracking is removed. If point sources in an image produce symmetric profiles, we can search for deviations in the skew or kurtosis of target asteroid profiles as evidence of an unresolved companion. Instead, we compare the profiles to models produced from field stars in the same image using a goodness-of-fit test, which is more robust against shared asymmetries.

In order to produce a cross-track profile of a target asteroid, we first take an aperture centered on the asteroid with size equal to 3 times the seeing. We upscale the aperture via block-replication (i.e., duplicating each pixel from the original aperture into a larger block of pixels) by a factor of 5, and rotate it using a third order spline interpolation, such that the direction of its non-sidereal motion falls along the x-axis. Then, we find the median of each row of pixels along the x-axis, producing the marginalized profile. Note that the median is more sensitive than the mean to a companion that is not positioned at right angle to their host's non-sidereal motion. Because different angles of rotation can cause a step-like pixelation in the resulting profile, we also use a 5 pixel moving average filter for smoothing.

For each image, we create a model profile by using the same process as for the target asteroid on at least 10 field stars, rotated to the same angle, and take their median and standard deviation. We reduce the impact of outliers by checking the goodness-of-fit of each field star against the median and clipping at the 90\% quantile. Each model profile is then normalized and centered on their respective target's cross-track profile.

The $\chi^{2}$ goodness-of-fit statistic between the target and reference profiles indicates the magnitude of the anomalous profile:

$$ \chi^{2}=\frac{1}{n}{\sum_{i=1}^{n}}\frac{\left(x_{i}-\overline{x}_{i}\right)^{2}}{\sigma_{i}^{2}} $$

The higher the statistic, the larger the deviation between the target and what would be expected for single asteroids, and therefore the possibility of an unresolved companion.

\section{Testing and Results}
In order to demonstrate proof of concept and to fine tune our methodology, it would be ideal to have a known binary asteroid that satisfies the following conditions:

\begin{enumerate}
  \item A primary that is bright but not saturated in observations ($m_V = 14-16$ for PS1 imagery), in order to reduce the impact of photon noise.
  \item A companion with a separation and contrast ratio within expected technique sensitivity range.
  \item A binary separation equal to or greater than the non-sidereal trailing within a single exposure so that the trailing does not dominate PSF orientation measurements.
  \item The companion magnitude contrast ratio is $\lesssim$ 3 magnitudes, as we expect this to be the limit extrapolated from the similar technique used by \citet{Deacon2017} on PS1 imagery.
\end{enumerate}

While there are currently no known main belt binary asteroids that meet all four criteria, we have no reason to expect fully qualified binaries to be particularly rare. Known binary asteroids easily fit three of the four criteria, e.g., 4674 Pauling reaches a brightness of $m_V\sim15.3$, with a companion at $\sim0\farcs4$ separation (0.4 Hill radius) and contrast ratio $\sim2.5$, but has non-sidereal trailing of $\sim0\farcs8$. The absence of a suitable target in the main belt may be due to the small number that have been observed by direct imaging.

Outside of the Main Belt, we applied the technique to 2577 Litva, which is a Mars-crossing asteroid that reaches a brightness of $m_V\sim14.2$ at opposition, with a companion with an estimated maximum separation of $\sim0\farcs7$ (0.73 Hill radius) and magnitude contrast ratio of $\sim2.6$. Under the 0.02 field star ellipticity threshold, two of nine images produced measurements unlikely to be a solitary asteroid. However, only one of the available images fell under the 0.01 field star ellipticity threshold, and that image did not produce a sufficiently anomalous measurement indicative of a possible companion.

Proceeding with application to our full target set in PS1 imagery, we searched the PS1 catalog (see \citealp{Flewelling2020}) for sources within 3\arcsec of each target's ephemeris positions. Sources flagged as ``known solar system object" or ``suspect object" were assumed to be detections of the target or transients. All other sources would contaminate our measurements, so where they were present and less than 5 magnitudes fainter than our target, those images were not used.

For the major axis orientation technique, we used ellipticity constraints of $<0.01$ and $<0.02$, requiring that at least 10 field stars in each cutout image satisfy the criteria. The first threshold corresponds to the median ellipticity of a simulated object under PS1 conditions with a companion at 1.5 pixel ($0\farcs39$) separation and contrast ratio of 3 magnitudes. As we expect this to be the limit of our sensitivity, PSF shape variances and/or shared asymmetries that produce ellipticities greater than this theshold will contaminate our measurements, resulting in false positives. This filtered out 94\% of the available imagery. The looser second threshold corresponds to the ellipticity \citet{Deacon2017} used to identify binary stars, and filtered out 73\% of the available imagery. Without filtering, the standard deviation of field star ellipticities can exceed $0.1$, suggesting that either PSF anisotropy for the specific image may be unusually high, or that some of the field stars are not singular point sources.

\begin{figure}
\includegraphics[scale=0.95]{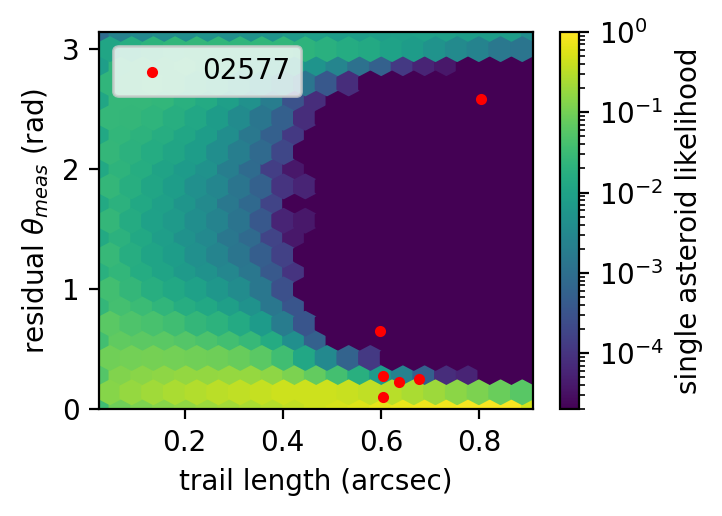}
\caption{Baseline hex-bin maps show the likelihood of a single asteroid resulting in the measured residual angle at the given non-sidereal trail length and observing conditions (here, PSF amplitude = 5000 DN, FWHM = 1\farcs5). Measurements that fall in low likelihood bins are possible indicators of asteroid binarity. Mars-crossing asteroid 2577 Litva, which has a known companion that we expect to be able to detect, shows two such ``hits". \label{fig:02577}}
\end{figure}

We then measured the PSF major axis orientation of the target asteroid in each remaining image, and matched them against hex-bin baselines with matching FWHM and peak flux, e.g., fig. \ref{fig:02577}. Measurements that fall in bins where the bin count is below a set threshold is considered a ``hit", where the measurement is unlikely to be due to a solitary asteroid. We scored the targets by number of hits and by the ratio of hits to images. Two different thresholds, at 10\% of the average hex-bin count (corresponding to a 0.4\% likelihood that a single asteroid will be measured in that specific bin) and at a stricter 1\% (0.04\% likelihood) of the average hex-bin count, provide us with binary asteroid candidate lists ranked by probability of having detected a companion (tab. \ref{tab:major_axis_candidates}).

\begin{figure}
\includegraphics[scale=0.95]{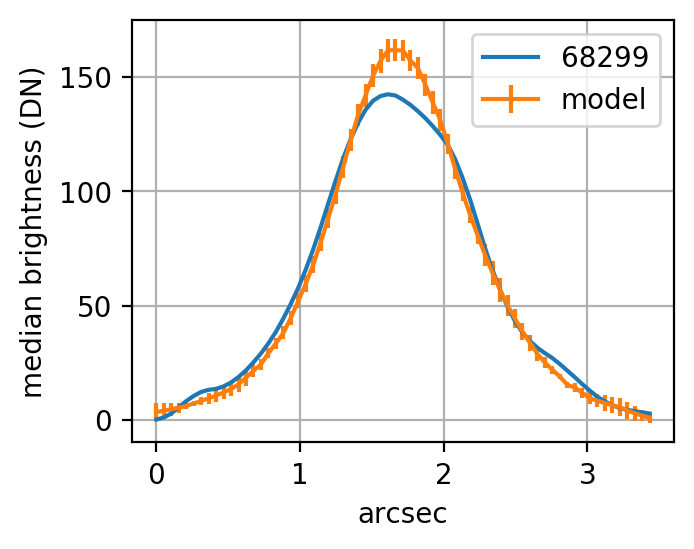}
\caption{This cross-track profile of asteroid 68299 is compared to the model profile created from field stars in the same image. The resulting chi-squared statistic is at the 95.1\% quantile, contributing to the asteroid's ranking as a binary candidate. \label{fig:68299}}
\end{figure}

\begin{table}
\caption{PSF Major Axis Orientation Binary Asteroid Candidates using 0.01 and 0.02 field star ellipticity constraints, and with 1\% and 10\% baseline bin count thresholds. \label{tab:major_axis_candidates}}
\begin{tabular}{|c|c|c|c|c|c|c|c|c|c|c|c|}
\hline 
\multicolumn{6}{|c|}{0.01 Field Star Ellipticity} & \multicolumn{6}{c|}{0.02 Field Star Ellipticity}\tabularnewline
\hline 
\multicolumn{3}{|c|}{1\% Bin Threshold} & \multicolumn{3}{c|}{10\% Bin Threshold} & \multicolumn{3}{c|}{1\% Bin Threshold} & \multicolumn{3}{c|}{10\% Bin Threshold}\tabularnewline
\hline 
Name & Hits & Ratio & Name & Hits & Ratio & Name & Hits & Ratio & Name & Hits & Ratio\tabularnewline
\hline 
\hline 
11841 & 5 & 1.00 &  2369 & 6 & 0.75 &  5191 & 19 & 0.39 &  5191 & 22 & 0.45\tabularnewline
\hline 
 3249 & 5 & 0.42 &  3973 & 6 & 0.75 &  4223 & 13 & 0.43 &  2808 & 16 & 0.36\tabularnewline
\hline 
 5191 & 4 & 0.40 & 11841 & 5 & 1.00 & 17814 & 12 & 0.29 &  4223 & 15 & 0.50\tabularnewline
\hline 
21787 & 4 & 0.36 &  1273 & 5 & 0.83 &  2740 &  9 & 0.28 & 17814 & 15 & 0.36\tabularnewline
\hline 
17860 & 3 & 0.60 &  3249 & 5 & 0.42 & 17860 &  8 & 0.38 &  2740 & 11 & 0.34\tabularnewline
\hline 
\end{tabular}
\tablecomments{Only the top five candidates under each ellipticity constraint and bin count threshold are shown here. These candidate lists are published in their entirety in the machine-readable format.}
\end{table}

For the cross-track profile technique, we produced the chi-squared goodness-of-fit statistic between the target and model profile (e.g., fig. \ref{fig:68299}) from each image. We then took the quantile at selected thresholds, and counted the number of images for each target with statistics above those quantiles. Manual examination of images beyond the 99\% quantile found that many were the result of either imaging artifacts or were affected by particularly bright objects outside our 3\arcsec contaminants search. Images in the 90\%-99\% range and in the 95\%-99\% range provide two candidate lists, with the top five of each list shown in table \ref{tab:crosstrack_candidates}.

\begin{table}
\caption{Cross-track Profile Binary Asteroid Candidates at 90\% and 95\% quantile thresholds. \label{tab:crosstrack_candidates}}
\begin{tabular}{|c|c|c|c|c|c|}
\hline 
\multicolumn{3}{|c|}{90\% Quantile} & \multicolumn{3}{c|}{95\% Quantile}\tabularnewline
\hline 
Name & Hits & Ratio & Name & Hits & Ratio\tabularnewline
\hline 
\hline 
16870 & 29 & 0.38 & 16870 & 16 & 0.21\tabularnewline
\hline 
21498 & 27 & 0.08 & 21498 & 16 & 0.05\tabularnewline
\hline 
12644 & 25 & 0.41 & 68299 & 15 & 0.38\tabularnewline
\hline 
 8724 & 24 & 0.39 & 36408 & 15 & 0.25\tabularnewline
\hline 
14480 & 24 & 0.38 &  9934 & 13 & 0.27\tabularnewline
\hline 
\end{tabular}
\tablecomments{Only the top five candidates above each quantile are shown here. These candidate lists are published in their entirety in the machine-readable format.}
\end{table}

We attempted to follow-up on some of our binary candidates using Keck II, but weather losses severely limited the number and priority of asteroids we were able to observe. The NIRC2 infrared camera behind the Keck II AO system has often been used to confirm the presence of companions detected by other telescopes (e.g., \citealp{Law2014}). Its narrow field camera has a pixel scale of 0.01 arcsec and has a 10x10 arcsec field of view. Using the $K_{short}$ filter, the diffraction limit is $\sim0\farcs05$.

With Keck II/NIRC2, we were able to observe three asteroids, 06441, 01419, and 29909, which were unfortunately not at the top of our priority lists. For each, we did not find a companion. Our companion detection limits (table \ref{tab:keck}), extrapolated from annulus noise measurements, show that it is very unlikely for Keck II/NIRC2 to fail to detect a real companion at the separations and contrast ratios where we expect the PSF major axis orientation and cross-track profile techniques to be sensitive.

\begin{table}
\caption{Keck II/NIRC2 companion detection limits in follow-up observations of selected targets.\label{tab:keck}}

\begin{tabular}{|c|c|c|c|c|c|c|}
\hline 
\multirow{2}{*}{Target} & \multirow{2}{*}{UT Date} & \multirow{2}{*}{FWHM} & \multirow{2}{*}{$N_{frames}$} & \multicolumn{3}{c|}{Limit (mag) at Sep.}\tabularnewline
\cline{5-7} \cline{6-7} \cline{7-7} 
 &  &  &  & 0.25'' & 0.50'' & 0.75''\tabularnewline
\hline 
\hline 
06441 & 20210225 & 0.14'' & 6 & 3.0 & 3.7 & 3.7\tabularnewline
\hline 
01419 & 20210402 & 0.06'' & 6 & 5.4 & 6.9 & 7.2\tabularnewline
\hline 
29909 & 20210402 & 0.07'' & 6 & 4.8 & 5.8 & 6.0\tabularnewline
\hline 
\end{tabular}
\end{table}

\section{Discussion}
Our simulations show that for solitary main belt asteroids, PSF major axis orientation measurements from the PS1 images should strongly agree with the direction of trailing produced by their non-sidereal motion, even where trailing lengths are $\sim1/3$ of the seeing. We suggest that any measurements that do not correlate with their non-sidereal rates are the result of either a companion asteroid or some other effect. Some of these possible effects are the contamination of the target field by background sources and systematic errors due to PSF anisotropy. While we worked to mitigate these effects by searching the PS1 stacked objects catalog for static objects and using only images where other point sources near the target appear sufficiently round, it is clear that our current processes have limitations. 

When cross-matching target lists, the PS1 stacked object search interface has a maximum search radius of 3 arcsec for each position. As PS1 has median seeing of 1\farcs0 - 1\farcs3, depending on filter, sufficiently bright objects just beyond the search radius can skew our measurements. We manually identified a few instances where this occurred when examining the results of the cross-track profile technique, but a more systematic method is required to handle the volume of imagery available. This also fails to take into account other non-sidereal objects or transients, though such events are very rare and unlikely to affect results.

\citet{Hoekstra1998} showed that the anisotropy of PSFs can depend on chip position by analyzing observations of globular clusters with the Hubble Space Telescope, and \citet{Magnier2018} showed how charge diffusion varies across the PS1 detector. Anisotropy may also be introduced by mechanical conditions, as well as in the transformation into ``warp" images for sky tessellation, such that the net effect may substantially differ between PS1 images. By not correcting for this effect and instead filtering out such a large fraction of images, it is possible that we have discarded the majority of observations where we may have a real chance at detecting a companion.

The probability that any given measurement is the result of a binary asteroid depends on the true separation and contrast distributions of companions; if a companion cannot have a stable orbit at a given separation, or if the only companions at that separation are small, then a measurement is unlikely to be the result of a companion, no matter how far a measurement deviates from that of our solitary asteroid baselines. With the sparse sampling of companions at wide separations, we consider the Hill radius as the outer bound, but the true region of stability for main belt binaries is likely smaller due to close approaches with other objects and orbital proximity to mean-motion or secular resonances.

\begin{figure}
\includegraphics[scale=0.95]{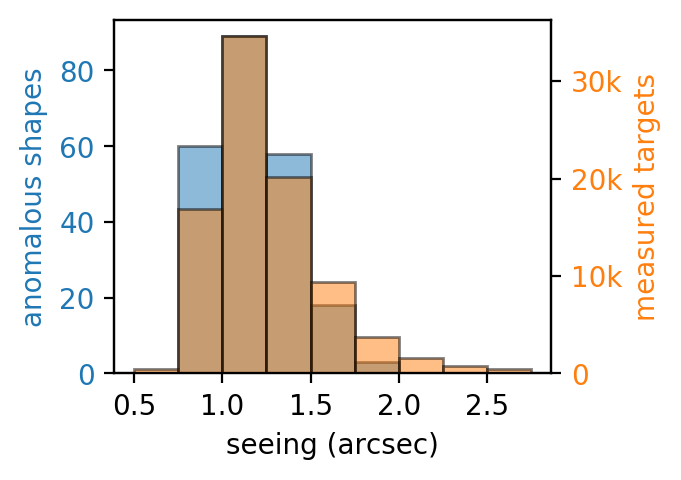}
\includegraphics[scale=0.95]{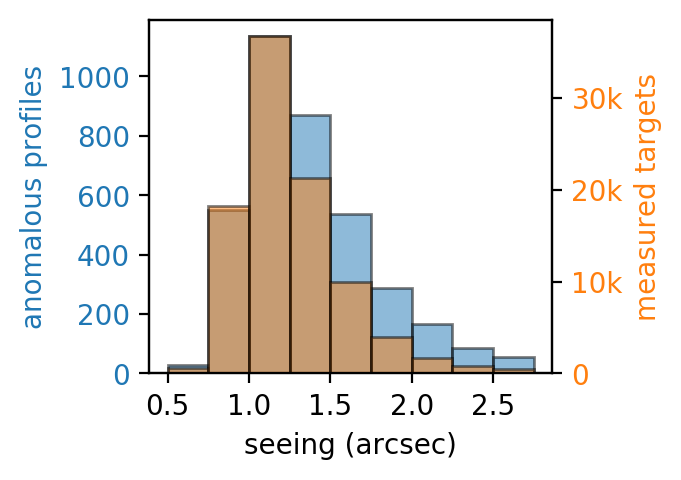}
\caption{The effects of seeing are captured in the production of the hex-bin maps in the shape measurement technique, so sensitivity to potential companions falls off at greater seeing (left). However, the presented cross-track profile technique does not compensate for noise, which likely produces more false-positives as seeing increases (right). \label{fig:sens_seeing}}
\end{figure}

\begin{figure}
\includegraphics[scale=0.95]{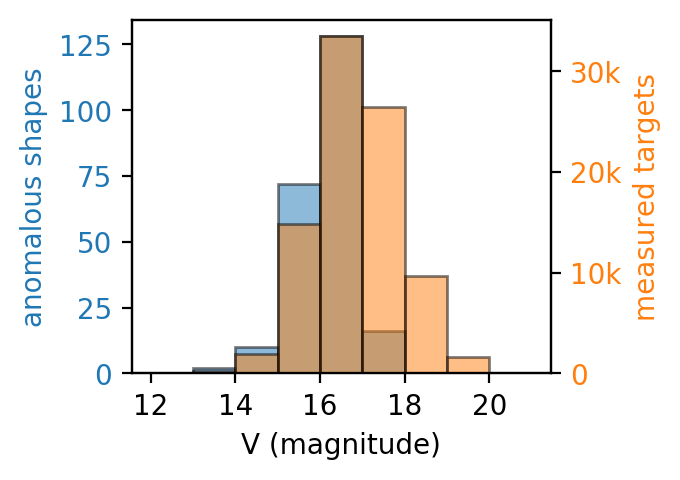}
\includegraphics[scale=0.95]{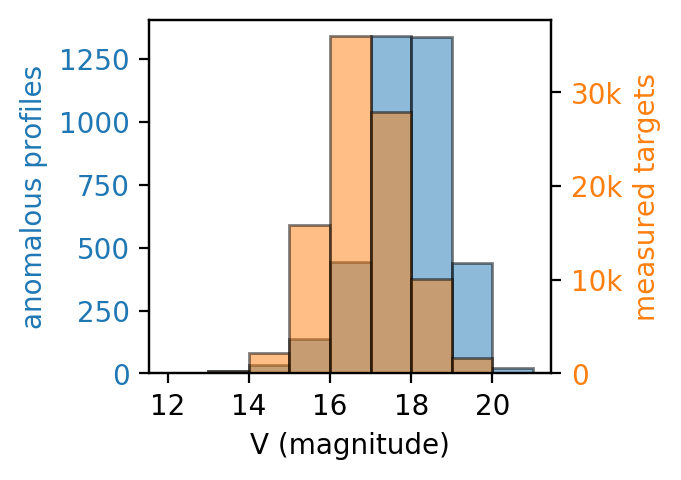}
\caption{Similar to the effects of seeing, shape measurement sensitivity falls off as targets become dimmer (left) due to its inclusion in the production of the hex-bin maps, whereas the cross-track profiles likely produces more false-positives due to the increase in relative noise (right). \label{fig:sens_vmag}}
\end{figure}

Examining the distributions of observing properties (seeing, brightness, non-sidereal trailing length, and trailing direction) for the anomalous measurements that may indicate the presence of a companion provides insight into our systematic biases. It is understood that the PanSTARRS survey may itself include observational selection effects, such as those described by \citet{Jedicke2002}, so it is necessary to compare against the distributions of the measured data set. As the seeing FWHM increases, the associated hex-bin heat maps used in the shape measurement technique have fewer available bins that fall under the outlier thresholds, and so is less likely to find higher seeing images to be anomalous. In contrast, the cross-track profiles are more likely to be divergent at higher seeings due to pixel noise and thus more likely to generate false-positives (fig. \ref{fig:sens_seeing}). The same behavior is seen for brightness, and for the same reason (fig. \ref{fig:sens_vmag}).

\begin{figure}
\includegraphics[scale=0.95]{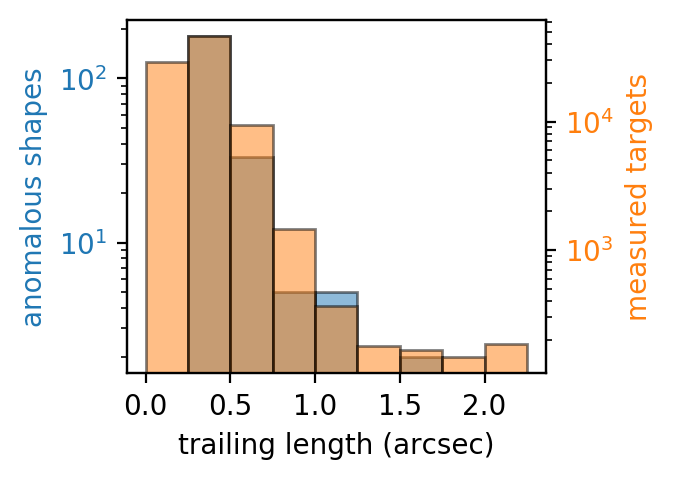}
\includegraphics[scale=0.95]{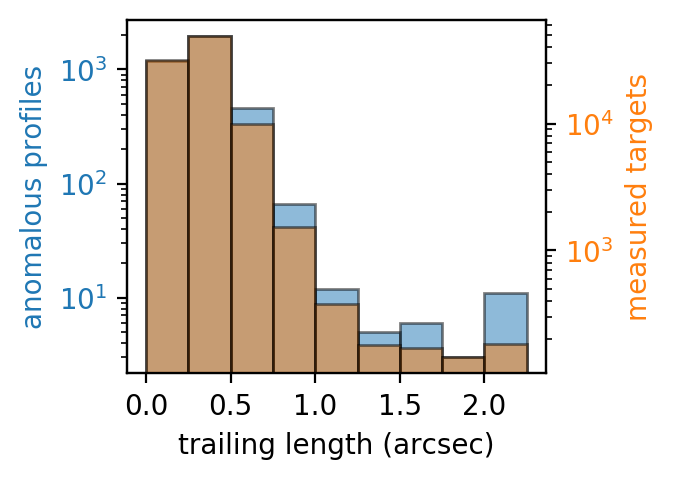}
\caption{The shape measurement technique requires that the non-sidereal trailing is neither too small nor too large; too short of a trail makes it difficult to differentiate anomalous measurements from those that can reasonably be attributed to noise, and trailing that is too long will dominate measurements such that any anomalies cannot be the result of a real companion within the host's Hill radius (left). The cross-track profiles are designed to minimize the effects of the trail, though there does appear to be a heightened sensitivity where trails are longer (right). \label{fig:sens_trail_len}}
\end{figure}

The shape measuring technique requires a certain minimum trailing length, depending on observing conditions, in order to sufficiently limit the expected measurements for solitary asteroids; this technique is not well suited to cases where the non-sidereal rate is too small or the integration time is too short, so the feasibility of the technique is limited by a combination of target brightness, heliocentric distance, and telescope size. Sensitivity also fails off for longer trails as greater length will dominate measurements. Cross-track profiles are designed to minimize the influence of the trailing on comparisons with image models, though the distributions seem to indicate a greater sensitivity at higher trailing lengths (fig. \ref{fig:sens_trail_len}). If this is real and not an increase in the false-positive rate, then it may be worth exploring the use of this technique on the faster moving inner solar system asteroid population.

\begin{figure}
\includegraphics[scale=0.95]{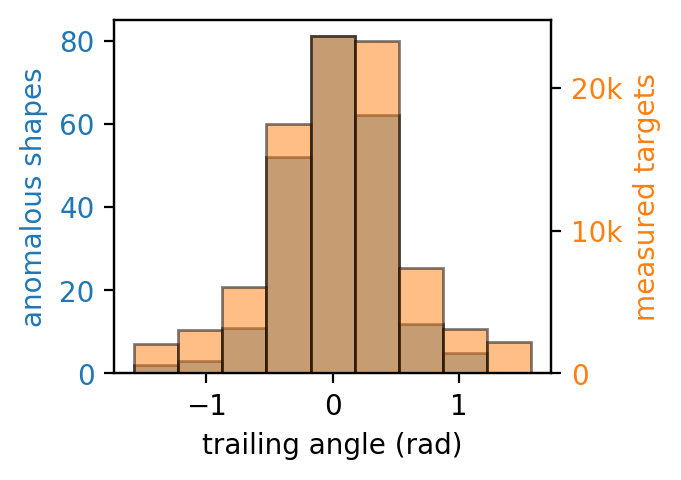}
\includegraphics[scale=0.95]{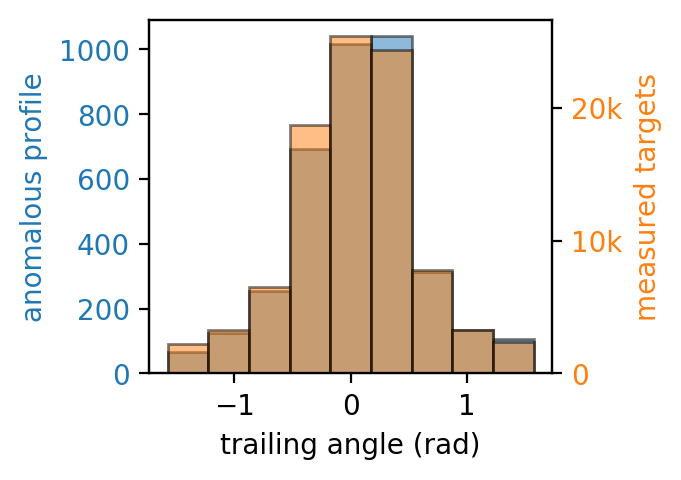}
\caption{The shape measurement technique appears to favor cases where the trailing angle is closely aligned with an image's x-axis (left), with nearly complete insensitivity as the angle approaches the y-axis (left). Cross-track profiles do not appear to have such a bias (right). \label{fig:sens_trail_theta}}
\end{figure}

While figure \ref{fig:meas-components} showed that the errors of angular measurements on square pixel data on the shape measuring technique is only significant at certain angles, figure \ref{fig:sens_trail_theta}) suggests that it may be enough to cause selection effects over a broader range of angles, with greater sensitivity where the trailing is closely aligned with the x-axis and/or less sensitive everywhere else. In contrast, the distributions comparison has not revealed an angular bias from the image rotation and measurement of cross-track profiles.

An ideal result would be to combine both techniques to produce a more robust candidate list. But examining the high scoring targets on each list have not identified clear points of agreement. It may be best to first (a.) optimize the techniques' thresholds against real binaries within technique sensitivity ranges, and (b.) develop a better understanding of false-positive rates, before attempting a combined solution. Both of these steps require substantially more follow-up with AO telescopes than we have been able to accomplish thus far.

\section{Conclusions and Future Work}
The PSF major axis orientation and cross-track profile techniques offer opportunities to increase the number of known binary asteroids, thus the sample size of asteroids with well determined masses, and in turn improve constraints on asteroid formation and evolution models. By using large scale, wide field survey imagery to identify candidates, we can potentially reduce the observational overheads needed to discover binary asteroids.

While we targeted members of main belt asteroid families with PS1 in this study, these techniques may be viable for other asteroid populations and/or other wide field surveys. Some refinement may be necessary as the populations will have differences in both trailing rates and potential angular separations. For example, the LSST on main belt targets, in addition to producing insufficient trailing for the major axis orientation technique, saturates at two magnitudes dimmer, corresponding to Hill radii 2.5 times smaller. This means that separations for which the LSST could be sensitive may fall outside the realm of stable orbits.

With enough images at a rapid, sustained cadence, it should be possible to greatly improve the robustness of our techniques. In particular, the relatively short orbital period of asteroid companions should correspond to cyclical behavior in our measurements. While there are challenges to the use of individual images from LSST, their baseline survey cadence would be excellent for this purpose. We can also expect an improvement in our analyses with Pan-STARRS as their continued operation produces a more numerous set of images to analyze.

To assess the effectiveness of both the PSF major axis orientation and cross-track profile techniques, higher angular resolution imaging is needed of the candidate asteroid binaries. If candidates are not confirmed, then knowledge of real binaries in the target parameter space will be required to enable refinement of the techniques.

This paper presents the first of our efforts to improve on our knowledge of binary asteroids in main belt asteroid families, and is limited to finding bright companions that are within 3 magnitudes (i.e., have diameters $\gtrsim$1/4) of their respective hosts. In order to search for binaries with greater contrast ratios, as well as to validate and improve on the techniques shown here by examining targets both on and off our candidate lists, we will be conducting a direct imaging survey with the upcoming Robo-AO-2 laser adaptive optics system on the UH 2.2-m telescope (\citealp{Baranec2018, Baranec2014}). We will be targeting several thousand members of the Vesta, Flora, and Eunomia asteroid families, and expect to find $>200$ companions based on the property distributions of the currently known binary population. This will enable statistical analyses of masses, porosities, and binary distributions on a per-family basis, and thus constrain theories of asteroid family formation.

\acknowledgments
The Pan-STARRS1 Surveys (PS1) and the PS1 public science archive have been made possible through contributions by the Institute for Astronomy, the University of Hawaii, the Pan-STARRS Project Office, the Max-Planck Society and its participating institutes, the Max Planck Institute for Astronomy, Heidelberg and the Max Planck Institute for Extraterrestrial Physics, Garching, The Johns Hopkins University, Durham University, the University of Edinburgh, the Queen's University Belfast, the Harvard-Smithsonian Center for Astrophysics, the Las Cumbres Observatory Global Telescope Network Incorporated, the National Central University of Taiwan, the Space Telescope Science Institute, the National Aeronautics and Space Administration under Grant No. NNX08AR22G issued through the Planetary Science Division of the NASA Science Mission Directorate, the National Science Foundation Grant No. AST-1238877, the University of Maryland, Eotvos Lorand University (ELTE), the Los Alamos National Laboratory, and the Gordon and Betty Moore Foundation.

This work has made use of the NASA/Jet Propulsion Laboratory (JPL) Horizons system (\url{https://ssd.jpl.nasa.gov/horizons/}). JPL Horizons provides access to key data and high accuracy ephemeris calculations of solar system objects.

JO and SJB acknowledge support from the NASA Infrared Telescope Facility, which is operated by the University of Hawaii under contract 80HQTR19D0030 with the National Aeronautics and Space Administration

We are grateful to Robert Weryk, who provided us with PS1 stamp request files for all of our targets, matching asteroid ephemerides to all available PS1 observations. We thank Michael Liu for assisting with Keck II observations, and Robert Jedicke for suggesting the cross-track profile technique.

\facilities{PS1 (GPC1), Keck:II (NIRC2)}

\bibliography{bib}
\bibliographystyle{aasjournal}

\end{document}